\documentclass[12pt]{article}
\usepackage{authblk}
\usepackage[bookmarksnumbered, colorlinks, plainpages]{hyperref}
\usepackage{amsmath, amsthm, amscd, amsfonts, amssymb, graphicx, color, booktabs}
\usepackage[square,numbers]{natbib}
\usepackage{caption, subcaption}
\usepackage{physics}
\numberwithin{equation}{section}
\definecolor{email}{rgb}{0.00,0.00,0.84}

\let\Im\undefined

\DeclareMathOperator{\Im}{Im}

\begin{document}
\setcounter{page}{1}

\title{\large \bf 12th Workshop on the CKM Unitarity Triangle\\ Santiago de Compostela, 18-22 September 2023 \\ \vspace{0.3cm}
\LARGE Model-independent description of $B\rightarrow D \pi \ell \nu$ decays}

\author[1]{Erik J. Gustafson}
\author[2]{Florian Herren}
\author[1]{Ruth S. Van de Water}
\author[3]{Raynette van Tonder}
\author[1]{Michael L. Wagman}
\affil[1]{Fermi National Accelerator Laboratory, Batavia,  Illinois, 60510, USA}
\affil[2]{Physics Institute, Universität Zürich,
Winterthurerstrasse 190, CH-8057 Zürich, Switzerland}
\affil[3]{Department of Physics, McGill University, 3600 rue University, Montréal, Québec, H3A 2T8, Canada}
\maketitle

\begin{abstract}
In this contribution we present a novel, model-independent description of semileptonic $B\rightarrow D \pi \ell \nu$ decays. In addition, we discuss recent developments in the understanding of coupled-channel $D \pi$-$D \eta$-$D_s K$ S-wave scattering and, for the first time, apply them to semileptonic decays. We not only obtain model-independent predictions for kinematic distributions in $B\rightarrow D \pi \ell \nu$ decays, but also rule out the hypothesis that the gap between the inclusive $B\rightarrow X\ell\nu$ branching fraction and the sum over exclusive channels is made up predominantly by $B\rightarrow D^{(\ast)} \eta \ell \nu$ decays.\\

\end{abstract} \maketitle

\section{Introduction}
Semileptonic $B\rightarrow D \pi \ell \nu$ decays, not including on-shell $B\rightarrow D^\ast (\rightarrow D \pi) \ell \nu$ decays, make up approximately 5\% of all semileptonic $B$ meson decays. Not only are they a signal component in inclusive $B\rightarrow X_c \ell \nu$ and $B\rightarrow X \tau \nu$ decays, but they also constitute an important background for studies of $B\rightarrow D^{(\ast)} \ell \nu$ decays, as well as measurements of $R(D^{(\ast)})$. Consequently, they contribute to both sides of the $|V_{cb}|$ inclusive-exclusive discrepancy and are relevant to determine if there are effects beyond the Standard Model in $b\rightarrow c\tau\nu$ transitions. Yet, experimental studies and the theoretical understanding of $B\rightarrow D \pi \ell \nu$ decays are not as mature as of $B\rightarrow D^{(\ast)} \ell \nu$ decays.

Quark models predict the existence of two low-lying doublets of excited $D$-meson states decaying to $D^{(\ast)}\pi$. The first one contains a scalar, the $D_0^\ast$, decaying to $D\pi$ and an axial-vector, the $D_1'$, decaying to $D^\ast\pi$ through the S-wave. Both are expected two have a large width due to their S-wave nature. The second doublet contains two narrow states: one axial-vector, the $D_1$, and a tensor, the $D_2^\ast$, which is the only of the four states decaying to both final states. The semileptonic decays of $B$ mesons into these four states are most commonly described by the HQET-based Leibovich-Ligeti-Stewart-Wise (LLSW) parametrization \cite{Leibovich:1997em,Leibovich:1997tu}, connecting transitions of $B$ mesons into the respective doublet partners.

On the experimental side, the masses and widths of the narrow states have been measured at the sub-MeV level by the LHCb collaboration in nonleptonic $B\rightarrow D^{(\ast)}\pi\pi$ decays. Yet, the masses and widths of the two broad states have large uncertainties since they do not appear as clear peaks in invariant mass spectra. Furthermore, the only available background-subtracted differential spectra in $B\rightarrow D \pi \ell \nu$ decays have been measured by the Belle experiment more than 15 years ago \cite{Belle:2007uwr}.

These spectra, together with the nonleptonic $B\rightarrow D^{\ast\ast}(\rightarrow D^{(\ast)}\pi)\pi$ branching ratios are the experimental input entering the two most detailed studies of $B\rightarrow D^{\ast\ast} \ell \nu$ decays \cite{Bernlochner:2016bci,Bernlochner:2017jxt,LeYaouanc:2022dmc}. Inspired by Dalitz-plot analyses in nonleptonic decays, the more recent study includes, in addition to the $D^{\ast\ast}$ modes, a possible virtual $D^{\ast}$ component, i.e. does account for the fact, that a very narrow Breit-Wigner distribution has a tail that drops like $1/(p^2-M^2)^2$.

This treatment of the $D^\ast$ is supported by the most recent study of $B\rightarrow D \pi \ell \nu$ decays by Belle \cite{Belle:2022yzd}, where a falling component is required to fit the data and a smaller than expected $D_0^\ast$ signal is observed.

\section{A model-independent parameterization}
In Ref.~\cite{Gustafson:2023lrz} we introduce a form-factor decomposition inspired by the treatment of $B\rightarrow D\ell\nu$ and $B\rightarrow D^\ast\ell\nu$ decays by Boyd, Grinstein and Lebed (BGL) \cite{Boyd:1995cf,Boyd:1995sq,Boyd:1997kz}, but extended, for the first time, to allow for two hadrons in the final state and arbitrary angular momenta of the intermediate states. The BGL parameterization itself is model-independent, but implements unitarity constraints on the $q^2$-dependence of form factors in a rigorous way. Consequently, it has proven to be very successful in experimental studies and Lattice QCD calculations of $B\rightarrow D^{\ast}\ell\nu$ decays.

The key behind the extension to multi-hadron final states is a partial-wave decomposition of the $D\pi$ system. This approach is natural and widely used in the study of nonleptonic three-body decays, as all hadronic resonances have definite angular momentum, e.g. the $D_2^\ast$ only appears in the $D\pi$ D-wave, but not in the P- or S-wave.
Thus, each partial wave is described by four (two for the S-wave) $q^2$- and $M_{D\pi}^2$-dependent form factors. Formally, the unitarity bounds are derived by considering the three-hadron contributions to two-point functions of the weak current. While each partial wave contributes to a given bound, due to the partial-wave expansion, there are no cross-terms, resulting in diagonal bounds.

To obtain a practically useful parameterization taking into account the unitarity bounds, we observe that the weak $b\rightarrow c$ transition takes place at much smaller length scales than the residual strong interactions between the two final-state hadrons. Thus, we write each form factor as
\begin{align}
f^{(l)}(q^2,M_{D\pi}^2) = \hat{f}^{(l)}(q^2,M_{D\pi}^2)g^{(l)}(M_{D\pi}^2)~,\label{eq:fffactorize}
\end{align}
where the function $g^{(l)}$ is the same for all form factors of a given partial wave and encodes the effect of final state interactions in the $D\pi$ system, such as the appearance of resonances. The remainder of the form factor only mildly depends on $M_{D\pi}^2$ and thus can be approximated. For the case of a partial wave with a single Breit-Wigner resonance of mass $M_R$, we could write:
\begin{align}
\hat{f}^{(l)}(q^2,M_{D\pi}^2) \approx \tilde{f}^{(l)}(q^2) + (M_R^2-M_{D\pi}^2)\bar{f}^{(l)}(q^2) + \mathcal{O}((M_R^2-M_{D\pi}^2)^2)~.\label{eq:ffaprox}
\end{align}
Neglecting all higher order terms, the function $\tilde{f}^{(l)}(q^2)$ can be treated just as a regular form factor in the BGL parameterization with modified outer functions encoding the effect of $g^{(l)}(M_{D\pi}^2)$:
\begin{align}
    \tilde{f}_l(q^2) = \frac{1}{\phi^{(f)}_l(q^2)B_f(q^2)}\sum_{i=0}^\infty a^{(f)}_{li} z^i,
\end{align}
where $B_f$ is a Blaschke factor including subthreshold $B_c$ resonances,
\begin{align}
    z(q^2,q^2_0) = \frac{q^2_0-q^2}{(\sqrt{q^2_+ - q^2}+\sqrt{q^2_+ - q_0^2})^2}~.
\end{align}
and the unitarity bound
\begin{align}
    \sum_{i,l} |a^{(f)}_{li}|^2 < 1~.
\end{align}
Including the suppressed term $\bar{f}^{(l)}(q^2)$ would lead to terms mixing the expansion coefficients of $\tilde{f}^{(l)}(q^2)$ and $\bar{f}^{(l)}(q^2)$ and consequently to a non-diagonal unitarity bound.

\begin{figure}
    \begin{subfigure}[b]{0.49\textwidth}
    \includegraphics[width=\textwidth]{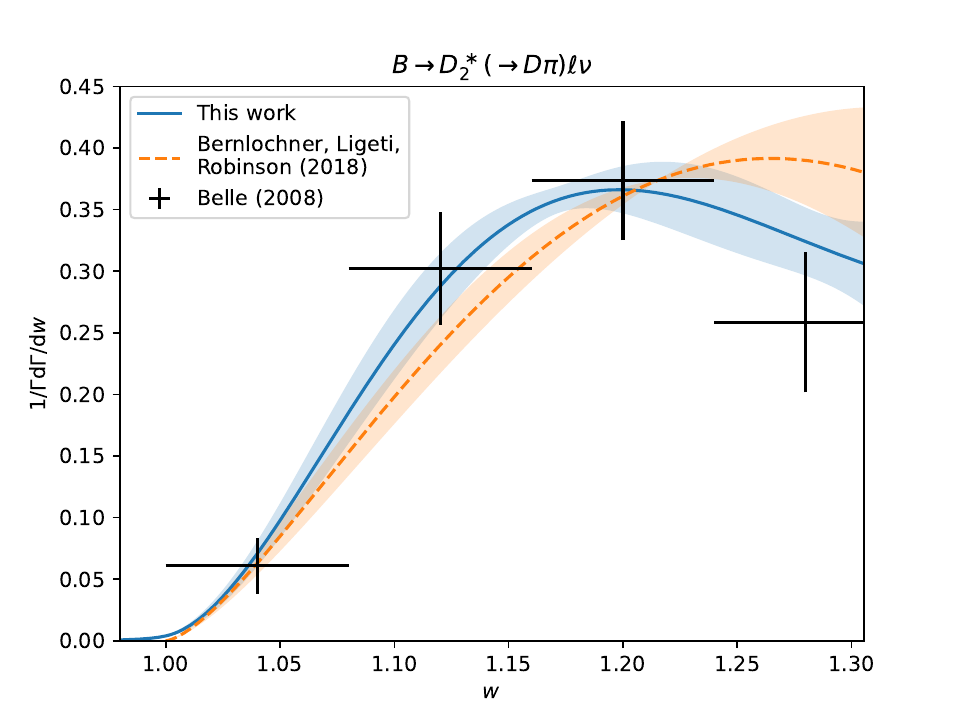}
    \caption{Normalized $B\rightarrow D_2^\ast \ell \nu$ $w$-spectrum}
    \label{fig:D2q2}
    \end{subfigure}
    \begin{subfigure}[b]{0.49\textwidth}
    \includegraphics[width=\textwidth]{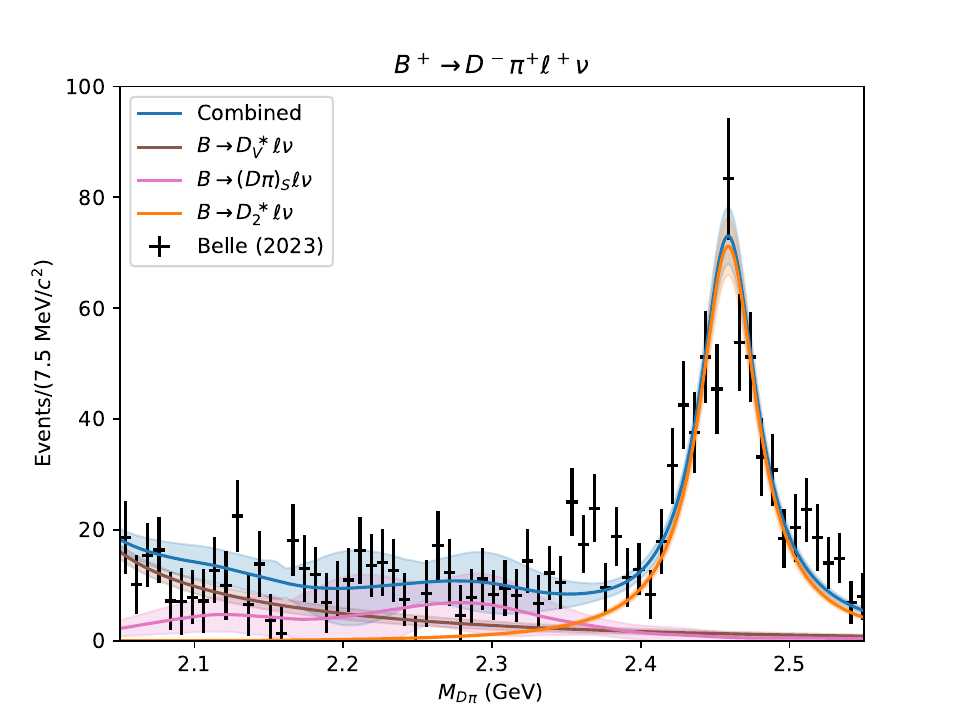}
    \caption{Fit of the measured $M_{D\pi}$-spectrum.}
    \label{fig:Mdpi}
    \end{subfigure}
\end{figure}

As a first application, we fit the $D\pi$ D-wave form factors to the differential decay rates measured by Belle \cite{Belle:2007uwr}. The resulting $w$-spectrum is shown in Fig.~\ref{fig:D2q2} and compared to the results of Ref.~\cite{Bernlochner:2017jxt}.

The second novelty of Ref.~\cite{Gustafson:2023lrz} is the treatment of the $D\pi$ S-wave contribution. Lattice QCD studies of $D\pi$ S-wave scattering \cite{Liu:2012zya,Moir:2016srx,Gayer:2021xzv} point to a lower mass of the $D_0^\ast$ than obtained from quark models: approximately $2.1$ GeV instead of $2.3$-$2.4$ GeV. In the context of unitarized chiral perturbation theory it was found that the calculation of Ref.~\cite{Liu:2012zya} leads to S-wave scattering matrices that contain two poles near $(2.1 - i 0.1)$ and $(2.45 - i 0.13)$ GeV~\cite{Albaladejo:2016lbb,Du:2017zvv,Guo:2017jvc}, with the former coupling predominantly to the $D\pi$ final state and the latter to the $D_s K$ final state. The resulting $D\pi$ lineshape can not be described in terms of a sum of Breit-Wigner curves and thus we follow a different strategy.

Below the onset of large $D\pi\pi\pi$ inelasticities, analyticity and unitarity dictate that the imaginary part of $f^{(0)}$ is given by the coupled-channel $D\pi$-$D\eta$-$D_s K$ scattering $T$-matrix, which we take from Ref.~\cite{Liu:2012zya}:
\begin{align}
\Im{\vec{f}(q^2, M_{D\pi}^2 + i\epsilon)} &= T^\ast(M_{D\pi}^2 + i\epsilon)\Sigma(M_{D\pi}^2)\vec{f}(q^2, M_{D\pi}^2 + i\epsilon)~,\label{eq:disphad}\\
\vec{f}(q^2,M_{D\pi}^2) &= \Omega(M_{D\pi}^2) \vec{P}(q^2,M_{D\pi}^2)~,\\
\Im{\Omega(s + i\epsilon)} &= \frac{1}{\pi}\int_{s_\mathrm{thr}}^\infty\frac{T^\ast(s')\Sigma(s')\Omega(s')}{s'-s-i\epsilon}\mathrm{d}s'~.
\end{align}
Here the vector $\vec{f}$ is a vector in channel-space, $\Sigma$ collects phase-space factor and $\Omega$ is the Muskhelishvili-Omnès matrix \cite{Omnes:1958hv,Muskhelishvili:1953}. The function $\vec{P}(q^2,M_{D\pi}^2)$ is a polynomial in $M_{D\pi}^2$ and we truncate it at zeroth order.

Combining our description of the S- and D-waves with the tail of the $D^\ast$ resonance in the P-wave, we fit to the $M_{D\pi}$ distributions recently measured by the Belle experiment \cite{Belle:2022yzd}. We obtain a good fit, the result displaced in Fig.~\ref{fig:Mdpi}, showing that semileptonic data are compatible with a two-pole structure in the S-wave.
However, in contrast to nonleptonic decays \cite{Du:2020pui} we can not rule out the quark model picture of a single, broad, S-wave resonance yet.

\section{Conclusion \& Outlook}
We have presented a model-independent parameterization of $B\rightarrow D\pi\ell\nu$ decay, a novel treatment of the $D\pi$ S-wave and compared to available data. The coupled channel treatment of the S-wave allows us to infer the branching ratios of $B\rightarrow D\eta\ell\nu$ and, through heavy quark spin-symmetry, $B\rightarrow D^\ast\eta\ell\nu$ decays, which are found to be at a level of $10^{-5}$. Thus, they can not account for the gap between the inclusive $B\rightarrow X\ell\nu$ branching fraction and the sum over exclusive states.

Our work opens the door to future studies of $1\rightarrow 2$-hadron semileptonic decays in a model-independent manner and will be crucial for direct measurements of the $D\pi$ S-wave scattering phase-shift, allowing to obtain the position of the lowest scalar $D$ meson pole from experiment.

\section*{Acknowledgements}
Fermilab is operated by Fermi Research Alliance, LLC under contract number DE-AC02-07CH11359 with the United States Department of Energy. FH acknowledges support by the Alexander von Humboldt foundation. RvT acknowledges support by the Natural Sciences and Engineering Research Council of Canada.
This research was supported in part by the Swiss National Science Foundation (SNF) under contract 200021-212729.
%---------------------------------------------------------------------------------------%
\bibliography{refs}

\end{document}